\documentclass[aps,prd,preprint,groupedaddress]{revtex4-1}

 \usepackage{graphicx,amsmath,amssymb}
 \usepackage{epstopdf}
\usepackage{lineno}
 \usepackage[usenames]{color}
 \usepackage[colorlinks=true, urlcolor=navyblue, linkcolor=navyblue, citecolor=navyblue]{hyperref}

\definecolor{navyblue}{rgb}{0,0.08,0.45}
\definecolor{darkred}{rgb}{0.7,0.0,0.0}
\definecolor{darkgreen}{rgb}{0,0.6,0.2}

\def\Dslash{\raise.15ex\hbox{/}\kern-.7em D}
\def\Pslash{\raise.15ex\hbox{/}\kern-.7em P}

\newcommand{\beq}{\begin{equation}}
\newcommand{\enq}{\end{equation}}
\newcommand{\beqa}{\begin{eqnarray}}
\newcommand{\beqast}{\begin{eqnarray*}}
\newcommand{\enqa}{\end{eqnarray}}
\newcommand{\enqast}{\end{eqnarray*}}
\newcommand{\nn}{\nonumber}
\newcommand{\req}[1]{(\ref{#1})}

\newcommand{\mbf}[1]{\mathbf{#1}}

\newcommand{\half}{{\frac{1}{2}}}
 \newcommand{\threehalf}{{\frac{3}{2}}}
 \newcommand{\fivehalf}{{\frac{5}{2}}}
 \newcommand{\sevenhalf}{{\frac{7}{2}}}
 \newcommand{\ninehalf}{{\frac{9}{2}}}
 \newcommand{\elevenhalf}{{\frac{11}{2}}}
 \newcommand{\thirteenhalf}{{\frac{13}{2}}}

\newcommand{\bec}{\begin{center}}
\newcommand{\enc}{\end{center}}
\newcommand{\beqo}{\begin{quote}}
\newcommand{\enqo}{\end{quote}}

\newcommand{\al}{\alpha}
\newcommand{\be}{\beta}
\newcommand{\ga}{\gamma}

\newcommand{\ep}{\epsilon}
\newcommand{\ze}{\zeta}

\newcommand{\ka}{\kappa}
\newcommand{\la}{\lambda}

\newcommand{\si}{\sigma}

\newcommand{\vp}{\varphi}

\newcommand{\Ga}{\Gamma}

\begin{document}

\preprint{SLAC--PUB--16130}

\title{Baryon Spectrum from Superconformal Quantum Mechanics and its Light-Front Holographic Embedding}

\author{Guy F.  de T\'eramond}
\affiliation{Universidad de Costa Rica, San Jos\'e, Costa Rica}
\email[]{gdt@asterix.crnet.cr}

\author{Hans G\"unter Dosch}
\affiliation{Institut f\"ur Theoretische Physik, Philosophenweg 16, D-6900 Heidelberg, Germany}
\email[]{h.g.dosch@thphys.uni-heidelberg.de}
\author{Stanley J. Brodsky}
\affiliation{SLAC National Accelerator Laboratory, Stanford University, Stanford, California 94309, USA}
\email{sjbth@slac.stanford.edu}

\date{\today}

\begin{abstract}

We describe the observed light-baryon spectrum by extending  superconformal quantum mechanics to the light front and its embedding in AdS space.  This procedure uniquely determines the confinement potential for arbitrary half-integer spin. To this end, we show that  fermionic wave equations in AdS space are dual to  light-front supersymmetric quantum mechanical bound-state equations in physical space-time. The specific breaking of conformal invariance  explains hadronic properties common to light mesons and baryons,  such as the observed mass pattern in the radial and orbital excitations, from the spectrum generating algebra. The holographic embedding in AdS also explains distinctive and systematic features, such as the spin-$J$ degeneracy for states with the same orbital angular momentum,  observed in the light baryon spectrum.

\end{abstract}


\maketitle


\section{Introduction}

The classical Lagrangian of QCD is invariant under scale and conformal transformations in the limit of massless quarks~\cite{Parisi:1972zy, Braun:2003rp}.  However, meson and baryon bound-states have well-defined ground states and towers of excited states with well defined and measurable properties such as mass and spin.  A simple but fundamental question in hadron physics is thus to understand the mechanism which endows a nominally conformal theory with a  mass scale, as well as to explain the remarkably similar linear Regge spectroscopy of both mesons and baryons.

In the quest for semiclassical equations to describe bound states, in the large distance strongly coupled regime of QCD, one can start by reducing the strongly correlated multi-parton light-front Hamiltonian dynamical problem to an effective one-dimensional quantum field theory~\cite{deTeramond:2008ht}.  This procedure is frame-independent and leads to a semiclassical, relativistic light-front (LF) wave equation for the valence state (the lowest Fock state), analogous to the Schr\"odinger  and Dirac equations in atomic physics.  The complexities arising from the strong interaction dynamics of QCD and an infinite class of Fock components  are incorporated in an effective potential $U$,  but its determination from first principles remains largely an open question.

Thus, a second central problem in the theoretical search for a semiclassical approximation to QCD is the construction of the effective LF confining potential $U$ which captures the underlying dynamics responsible for confinement, the emergence of a mass scale as well as the universal Regge behavior of mesons and baryons. Since our light-front semiclassical approach is based on a one-dimensional quantum field theory, it is natural to extend the framework introduced  by V. de Alfaro, S. Fubini and G. Furlan (dAFF)~\cite{deAlfaro:1976je}  to the frame-independent light-front Hamiltonian theory, since it gives important insight into the QCD confinement mechanism~\cite{Brodsky:2013ar}. Remarkably, dAFF show that a mass scale can appear in the Hamiltonian without breaking the conformal invariance of the action.

The dAFF construction~\cite{deAlfaro:1976je} begins with the study of the spectrum of a conformally invariant one-dimensional quantum field theory which does not have a normalizable ground state. A new Hamiltonian is defined as a superposition of the generators of the conformal group and consequently it leads to a redefinition of the corresponding evolution parameter $\tau$, the range of which is finite. This choice determines the quantum mechanical evolution of the system in terms of a compact operator with normalizable eigenstates and a well defined ground state.    A scale appears in the Hamiltonian  while retaining the conformal invariance of the action~\cite{deAlfaro:1976je}.  This remarkable result is based on the isomorphism of the algebra of the one-dimensional conformal group {\it Conf}$(R^1)$ to the algebra of generators of the group $SO(2,1)$.  One of the generators of this group, the rotation in the 2-dimensional space, is compact. As a result, the form of the evolution operator  is fixed and includes a confining harmonic oscillator potential,  thus equally spaced eigenvalues~\cite{Leutwyler:1977pv, Trawinski:2014msa}.  Since the generators of {\it Conf}$(R^1)$ have different dimensions, their relations with the generators of $SO(2,1)$ imply a scale, which according to dAFF may play a fundamental role~\cite{deAlfaro:1976je, Brodsky:2013ar}.

A third important feature in the construction of  semiclassical equations in QCD, is the correspondence between the equations of motion  for arbitrary spin in  Anti--de Sitter (AdS) space and the light-front  Hamiltonian equations of motion for relativistic light hadron bound-states in physical space-time~\cite{deTeramond:2008ht, deTeramond:2013it}. This approach is inspired by the AdS/CFT correspondence~\cite{Maldacena:1997re} where, in principle, one can compute physical observables in a strongly coupled gauge theory in terms of a weakly coupled classical gravity theory defined in a higher dimensional space~\cite{Maldacena:1997re, Gubser:1998bc, Witten:1998qj}.  In fact, an additional motivation for using  AdS/CFT  ideas to describe strongly coupled QCD follows from the vanishing of the $\beta$-function in the infrared, which leads to a conformal window in this regime~\cite{footnote0, Brodsky:2010ur, Deur:2014qfa}.

The procedure, known as light-front holography~\cite{Brodsky:2006uqa, deTeramond:2008ht, Brodsky:2014yha}, allows one to establish a precise relation between wavefunctions in AdS space and the LF wavefunctions  describing the internal structure of hadrons.  As a result, the effective LF potential $U$ derived from the AdS embedding  is conveniently expressed,  for arbitrary integer spin representations, in terms of a dilaton profile which is determined by the dAFF procedure described above~\cite{deTeramond:2013it, Brodsky:2013ar}.  The result is a light-front wave equation which reproduces prominent aspects of hadronic data, such as the mass pattern observed in the radial and orbital excitations of the light mesons~\cite{Brodsky:2014yha}, and in particular a massless pion in the chiral limit.

The light-front holographic embedding for baryons  is not as simple as for mesons, since a dilaton term in the AdS fermionic action can be rotated away by a redefinition of the fermion fields in AdS~\cite{Brodsky:2014yha, Kirsch:2006he}, and therefore it has no dynamical effects on the spectrum.  In practice, one can introduce an effective interaction  in the fermion action, a Yukawa term,   which breaks the maximal symmetry in AdS and consequently the conformal symmetry in Minkowski space.  This leads to a linear confining interaction in a LF Dirac equation for baryons whose eigensolutions generate a baryonic Regge spectrum~\cite{Brodsky:2008pg, Abidin:2009hr}. The confining interaction term can be constrained by the condition that the ÔsquareÕ of the Dirac equation leads to a potential which matches the form of the dilaton-induced potential for integer spin, but this procedure appears to be {\it ad-hoc}.

There are some striking similarities between the spectra of the observed  light mesons and baryons: they are of similar mass, the slope and spacing of the quantum orbital excitations in $L$ and their daughter spacing in $n$, the radial quantum number, is the same.  This behavior in the meson sector is related to the introduction of a scale within the framework of the  conformal algebra. This procedure leaves the action invariant~\cite{deAlfaro:1976je, Brodsky:2013ar}.  Since supersymmetry is related with boson-fermion symmetry, it is compelling to examine the properties of the supersymmetric  algebra and its superconformal extension to describe baryons in complete analogy to the bosonic case, where the confining potential was determined by the conformal algebra of  one-dimensional quantum field theory~\cite{deAlfaro:1976je, Brodsky:2013ar}.  In fact, it is straightforward to translate a quantum mechanical model into its supersymmetric (SUSY) counterpart by following Witten's construction~\cite{Witten:1981nf}. Superconformal quantum mechanics, the supersymmetric extension~\cite{Akulov:1984uh, Fubini:1984hf} of conformal quantum mechanics~\cite{deAlfaro:1976je}, then follows from the properties of the  superconformal algebra.

We shall show in this article that the structure of supersymmetric quantum mechanics is encoded holographically in the AdS equations for arbitrary half-integer spin for any superpotential.  Most important for the present discussion, we will show that superconformal quantum mechanics~\cite{Fubini:1984hf}  has an elegant representation on the light front and its holographic embedding in AdS space. Remarkably, this procedure uniquely determines the form of the confinement potential for arbitrary half-integer spin.  If one extends with Fubini and Rabinovici~\cite{Fubini:1984hf}, the method of de Alfaro, Fubini and Furlan~\cite{deAlfaro:1976je}  to the superconformal algebra, the form of the potential in the new evolution equations is completely fixed.  We will also discuss in this article how the different embeddings of mesons and baryons in AdS space~\cite{deTeramond:2013it} lead to distinct systematic features of meson and baryon spectroscopy.  In particular, we will show that the integrability methods used to construct baryonic light-front equations~\cite{Brodsky:2008pg} are the light-front extension of the usual formulation of  supersymmetric  Hamiltonian quantum mechanics~\cite{Witten:1981nf, Cooper:1994eh}. In fact, a possible indication of a supersymmetric connection was already mentioned in Ref.~\cite{Brodsky:2008pg}, but a proof was not actually given there~\cite{Pallares-Rivera:2014mza}.

This article is organized as follows:  In Sec.  \ref{LFCQM} we review for convenience light-front conformal quantum mechanics and its holographic embedding in AdS space. In Sec. \ref{LFSUSYQM} we extend  supersymmetric  quantum mechanics to the light-front and describe its embedding in AdS space. We show in particular that properly taking the square root of the light-front Hamiltonian operator leads to a linear relativistic invariant Dirac equation.  In Sec. \ref{LFSCQM} superconformal quantum mechanics is extended to  light-front holographic QCD.  The application of the method  to the complex patterns observed in  baryon spectroscopy  is discussed in Sec. \ref{BS}. Some final comments and conclusions are given in Sec. \ref{conclusions}. In  Appendix  \ref{appendix} we  discuss briefly the specific action of the supercharges.

\section{Light-Front Conformal Quantum Mechanics  and its Holographic Embedding\label{LFCQM}}

Following Ref.~\cite{deAlfaro:1976je} one starts with the one-dimensional action
\beq \label{SQ}
S[x] = \half \int dt \left (\dot x^2 - \frac{g}{x^2} \right),
\enq
where  $x(t)$ is a  field operator, the constant $g$ is dimensionless, and $t$ has dimensions of length squared. The action \req{SQ} is invariant under conformal transformations in the variable $t$, thus in addition to the Hamiltonian  $H$  there are two more invariants of motion, namely the dilatation operator  $D$ and the operator of special conformal transformations $K$, corresponding to the generators of the conformal group {\it Conf}$(R^1)$ with commutation relations
 \beq  \label{CA}
 [H,D]= i\,H,  \quad [H ,K]=2\, i \, D, \quad [K,D]=- i\,  K .
\enq

Specifically, if one introduces the  new variable $\tau$ defined through the relation
\beq 
d\tau= \frac{d t}{u+v\,t + w\,t^2},
\enq
 it then follows that the operator 
 \beq
 G= u\,H+ v\,D + w\,K,
 \enq
 generates the quantum mechanical unitary evolution in  $\tau$~\cite{deAlfaro:1976je}
\beq
G \vert \psi(\tau) \rangle = i \frac {d}{d \tau} \vert \psi(\tau)\rangle.
\enq
One can show that $G$ is a compact operator  for $4 u w - v^2 > 0$~\cite{deAlfaro:1976je}.  In terms of the fields $x$ and $p = \dot x$  the new Hamiltonian $G$ is given by
\beq \label{HtauQ}   
G (x, p ) =  \frac{1}{2} u \left(p^2 + \frac{g}{x^2} \right)  - \frac{1}{4} v \left( x p + p x\right) + \frac{1}{2} w x^2,
 \enq
at $t = 0$. In the Schr\"odinger representation $x(0)$ is represented by the position operator and $p \to - i \frac{d}{d x}$. The Hamiltonian is~\cite{deAlfaro:1976je} 
\beqa \label{G} 
G \! &=& \!  {1\over 2} u \Big(-{d^2\over dx^2}  + {g\over x^2} \Big) + {i\over 4} v \Big(x {d\over dx} + {d\over dx}x \Big) + {1\over 2} w x^2, \\
  \! &= \! & u H + v D + w K,  \nn
\enqa
with
\beqa  
H  & \! =  \! &  {1\over 2} \Big(-{d^2\over dx^2}  + {g\over x^2} \Big) , \\
D  & \! = \! & {i\over 4}  \Big(x {d\over dx} + {d\over dx}x \Big),\\
K  &\! = \!  &{1\over 2} w x^2,
\enqa
the superposition of the `free' Hamiltonian $H$, the generator of dilatations $D$ and the generator of special conformal transformations $K$ in one dimension.

We now compare the dAFF Hamiltonian with the light-front Hamiltonian in the semiclassical approximation described in~\cite{deTeramond:2008ht}.  A physical hadron in four-dimensional Minkowski space has four-momentum $P_\mu$ and invariant hadronic mass squared $H_{LF} = P_\mu P^\mu = M^2$~\cite{Footnote1, Brodsky:1997de}. In the limit of zero quark masses the longitudinal modes decouple and the LF eigenvalue equation $H_{LF} \vert \phi \rangle  =  M^2 \vert \phi \rangle$ is a light-front  wave equation for $\phi$~\cite{deTeramond:2008ht}
\begin{equation} \label{LFWE}
\left(-\frac{d^2}{d\zeta^2}
- \frac{1 - 4L^2}{4\zeta^2} + U\left(\zeta, J\right) \right)
\phi(\zeta) = M^2 \phi(\zeta),
\end{equation}
a relativistic single-variable  LF  Schr\"odinger equation.  The boost-invariant transverse-impact variable $\zeta$~\cite{Brodsky:2006uqa} measures the separation of quark and gluons at equal light-front time~\cite{Footnote2}, and  it also allows one to separate the bound-state dynamics of the constituents from the kinematics of their LF internal angular momentum $L$  in the transverse light-front plane~\cite{deTeramond:2008ht}. The effective interaction $U$ is instantaneous in LF time and acts on the lowest state of the LF Hamiltonian. To actually compute $U$ in the semiclasscal approximation  one must systematically express higher Fock components as functionals of the lower ones. This method has the advantage that the Fock space is not truncated and the symmetries of the Lagrangian are preserved~\cite{Pauli:1998tf}.

Comparing the Hamiltonian $G$ \req{G}  with the light-front wave equation \req{LFWE} and identifying the variable $x$ with the light-front invariant variable $\zeta$,  we have to choose $u=2, \; v=0$ and relate the dimensionless constant $g$  to the LF orbital angular momentum, 
\beq  \label{gL}
g=L^2 - \frac{1}{4},
\enq
 in order to reproduce the light-front kinematics. Furthermore  $w = 2 \lambda_M^2$ fixes the form of the  confining light-front  potential  to that of a harmonic oscillator in the LF transverse plane~\cite{Brodsky:2013ar},  
 \beq \label{Uconst}
 U \sim \la_M^2 \, \zeta^2.
 \enq  
 In contrast to the baryonic case, which is discussed below, one can perform a constant  level shift by adding an arbitrary constant, with dimension mass squared, to the confining term in the light front potential.

\subsection{Light-Front Holographic Embedding \label{LFEM}}

The next step is to determine the arbitrary constant term in the LF effective potential for arbitrary integer spin representations. Following Ref.~\cite{deTeramond:2013it} this constant term in the potential is determined by the embedding of the LF Hamiltonian equations in AdS space. To this end it is convenient to consider an effective action for a spin-$J$ field in AdS$_{d+1}$ space represented by a totally symmetric rank-$J$ tensor field {  $\Phi_{N_1  \dots N_J}$, where $M,N$ are  the indices of the  $d + 1$ higher dimensional AdS space with coordinates $x^M =\left(x^\mu, z\right)$. The coordinate $z$ is the holographic variable and the $x^\mu$ are Minkowski flat space-time coordinates. In the presence of a dilaton background $\varphi$ the effective action in~\cite{deTeramond:2013it} is
\begin{multline}
\label{action2}
S_{\it eff} = \int d^{d} x \,dz \,\sqrt{g}  \; e^{\vp(z)} \,g^{N_1 N_1'} \cdots  g^{N_J N_J'}   \Big(  g^{M M'} D_M \Phi^*_{N_1 \dots N_J}\, D_{M'} \Phi_{N_1 ' \dots N_J'}  \\
 - \mu_{\it eff}^2(z)  \, \Phi^*_{N_1 \dots N_J} \, \Phi_{N_1 ' \dots N_J'} \Big),
 \end{multline}
where  $\sqrt{g} = (R/z)^{d+1}$ and $D_M$ is the covariant derivative which includes the affine connection ($R$ is the AdS radius).  The dilaton  $\varphi(z)$ effectively breaks the maximal symmetry of AdS, and the $z$ dependence of the effective AdS mass $\mu_{eff}$ allows a clear separation of kinematical and dynamical effects. It is determined  by the precise mapping of AdS to light-front physics~\cite{deTeramond:2013it}.

In order to map to the LF Hamiltonian, one considers hadronic states with momentum $P$ and a  $z$-independent spinor $\ep_{\nu_1 \cdots \nu_J}(P)$   with polarization components along the physical Minkowski coordinates. In holographic QCD such a state is described by the product of  a free state with moment $P$,  propagating in physical space-time,
and $z$-dependent wave function $\Phi_J$ 
\beq \label{scalarcov} \Phi_{\nu_1 \cdots \nu_J}(x, z) = 
e ^{ i P \cdot x} \,  \ep_{\nu_1 \cdots \nu_J}({P}) \,  \Phi_J(z) , \enq
with invariant hadron mass $P_\mu P^\mu \equiv \eta^{\mu \nu} P_\mu P_\nu = M^2$. Variation of the action leads to the wave equation
\beq  \label{PhiJM}
 \left[
   -  \frac{ z^{d-1- 2J}}{e^{\varphi(z)}}   \partial_z \left(\frac{e^{\varphi(z)}}{z^{d-1-2J}} \partial_z   \right)
  +  \frac{(\mu \,R )^2}{z^2}  \right]  \Phi_J = M^2 \Phi_J,
  \enq
where   $(\mu \, R)^2 = (\mu_{\it eff}(z) R)^2  - J z \, \vp'(z) + J(d - J +1)$ is a constant determined by kinematical conditions in the light front~\cite{deTeramond:2013it}.  Variation of the  AdS action also gives  the kinematical constraints required to eliminate the lower spin states $J-1, \,J-2, \cdots$ from the fully symmetric  AdS  tensor field $\Phi_{\nu_1 \dots \nu_J}$~\cite{deTeramond:2013it}:
\beq \label{sub-spin}
 \eta^{\mu \nu } P_\mu \,\ep_{\nu \nu_2 \cdots \nu_J}=0, \quad \quad
\eta^{\mu \nu } \,\ep_{\mu \nu \nu_3  \cdots \nu_J}=0.
 \enq

We now perform the AdS  mapping to  light-front physics in physical space-time.   To this end we factor out the scale  $(1/z)^{ J - (d-1)/2}$ and  dilaton factors from the AdS field writing
\beq
\Phi_J(z)   = \left(R/z\right)^{J - (d-1)/2 } e^{- \varphi(z)/2} \, \phi_J(z).
\enq
Upon the substitution of the holographic variable $z$ by the light-front invariant variable $\zeta$ and replacing $\Phi_J$  into the AdS eigenvalue equation  (\ref{PhiJM}), we obtain for $d=4$ the QCD light-front frame-independent wave equation \req{LFWE}  with the effective LF potential~\cite{deTeramond:2010ge, deTeramond:2013it}
\begin{equation} \label{U}
U(\zeta, J) = \half \varphi''(\zeta) +\frac{1}{4} \varphi'(\zeta)^2  + \frac{2J - 3}{2 \zeta} \varphi'(\zeta) .
\end{equation}
The fifth dimensional AdS mass $\mu$ in (\ref{PhiJM}) is related to the light-front internal orbital angular momentum $L$  and the total angular momentum $J$ of the hadron according to
\beq \label{muJL}
(\mu R)^2 = - (2-J)^2 + L^2,
\enq
where the critical value  $L=0$  corresponds to the lowest possible stable solution~\cite{Breitenlohner:1982jf}.

From the holographic relation \req{U} it follows that the harmonic potential  is holographically related to a unique dilaton profile, $\varphi = \lambda_M z^2$ provided that $\varphi(z) \to 0$ as $z \to 0$. From \req{U} we find the effective LF potential  \req{Uconst} 
\beq \label{UJ} 
U(\ze, J) =   \la_M^2 \ze^2 + 2 \la_M (J - 1).
\enq
The term  $\la_M^2 \ze^2$ is determined uniquely by the underlying conformal invariance of the one-dimensional effective theory, and the constant term $2 \la_M (J-1)$ is determined by the spin representations in the embedding space.

For the effective potential (\ref{UJ}) equation  (\ref{LFWE}) has eigenfunctions
\beq \label{phi}
\phi_{n, L}(\zeta) = \vert \la_M\vert^{(1+L)/2} \sqrt{\frac{2 n!}{(n\!+\!L\!)!}} \, \zeta^{1/2+L}
e^{- \vert \la_M \vert \zeta^2/2} L^L_n(\vert \la_M \vert \zeta^2) ,
\enq
and eigenvalues
\beq\label{M2SFM}
M_{n, J, L}^2 = 4 \la_M \left(n + \frac{J+L}{2} \right),
\enq
for $\la_M>0$. The spectral predictions  explain the essential features of the observed light meson spectrum~\cite{Brodsky:2014yha}, including a zero pion mass in the chiral limit, and Regge trajectories with the same slope in the quantum numbers $n$ and $L$. The solution for $\la_M<0$  leads to a meson spectrum in clear disagreement with observations. Since the effective interaction is determined from the  conformal symmetry of the effective one-dimensional quantum field theory, which is not severely broken for small quark masses, the method can be successfully extended to describe, for example, the $K$ and $K^*$ excitation spectrum~\cite{deTeramond:2014rsa, Brodsky:2014yha}.

\section{Light-Front Supersymmetric  Quantum Mechanics and its Holographic Embedding \label{LFSUSYQM}}
 
Supersymmetric quantum mechanics is a simple realization of a graded Lie algebra  which contains two fermionic generators, the supercharges, $Q$ and $Q^\dagger$, and a bosonic generator, the Hamiltonian $H$, which are operators in a state space~\cite{Witten:1981nf}. It closes under the graded algebra $sl(1/1)$:
\beqa
&&\half \{Q,Q^\dagger\} =   H,  \\
&&\{Q,Q\} =   \{Q^\dagger, Q^\dagger\} = 0, \\
&&{[Q, H]}  = [Q^\dagger, H] = 0. \label{QHcr}
\enqa

It is useful to write down the SUSY formulation of quantum mechanics in terms of  anti-commuting spinor operators $\chi$.  A minimal realization of the group generators is given in terms of the one-dimensional representation
\beq
 Q =    \chi \left(\frac{d}{d x} + W(x) \right) \label{Q} ,
 \enq
 and
 \beq
 Q^\dagger = \chi^\dagger \left(- \frac{d}{d x} + W(x)\right) \label{Qdag} ,
 \enq
where  $W(x)$ is called the superpotential in the context of supersymmetric theories.
The spinor operators $\chi$ and $\chi^\dagger$ satisfy the anti-commutation relation
 \beq
 \{\chi, \chi^\dagger\} = 1.
 \enq
 Using a representation in terms of $2 \times 2$ Pauli-spin matrices we have
\beq
\chi  = \frac{\si_1 + i  \si_2}{2},  \quad  \chi^{\dagger} = \frac{\si_1 - i \si_2}{2},
\enq
and
\beq  \label{si3}
 [\chi, \chi^\dagger] = \si_3.
\enq
Thus the Hamiltonian is
\beq
H = \half\{Q, Q^\dagger\} = \frac{1}{2} \left(- \frac{d^2}{d x^2}  + W^2(x) +   \si_3 \, W'(x) \right).
\enq
It can be written in matrix form:
\beq
H =   \frac{1}{2}
 \left( \begin{array}{cc}
    H_+ & 0\\ 
    0 & H_- \\  
  \end{array} \right) =
  \frac{1}{2}
 \left( \begin{array}{cc}
    - \frac{d^2}{d x^2} + V_+(x) & 0\\ 
    0 & - \frac{d^2}{d x^2} + V_-(x) \\  
  \end{array} \right),
  \enq
with effective potential 
\beq \label{HVpm}
V_\pm(x) = W^2(x) \pm W'(x).
\enq
Since $H$ commutes with $Q$ and $Q^\dagger$ \req{QHcr}, it follows that the eigenvalues of $H_+$ and $H_-$ are  identical.

\subsection{Supersymmetric Quantum Mechanics in the Light-Front \label{LFSUSYQMA}}

To give a relativistic formulation of  supersymmetric quantum mechanics  it is convenient to write the anti-commuting spinor operators in terms of a $4 \times 4$  matrix representation of the Clifford algebra, which acts on four-dimensional physical space where the LF spinors are defined.  We use the Weyl representation where the chirality operator  $\ga_5$ is diagonal,  and define the matrices $\al$ and $\be$ by
\beq 
 i \alpha =
  \begin{pmatrix}
  0& I\\
- I& 0
  \end{pmatrix}, ~~~~~
 \beta =
  \begin{pmatrix}
  0& I\\
  I& 0
  \end{pmatrix},~~~~~
\enq
where $I$ a two-dimensional unit matrix.  The matrices $\al$ and $\be$ are hermitian and anti-commuting
\begin{eqnarray} \label{DM1}
\alpha^\dagger = \alpha,~~~ \alpha^2 = 1,\\ 
\beta^\dagger = \beta,~~ \be^2 = 1,\\ 
\{\alpha, \beta\} = 0.
\end{eqnarray}
From the product of $\al$ and $\be$ we construct a third matrix $\ga_5$, which corresponds to the usual chirality operator: $\ga_5 = i \al \be$
\beq \label{ga}
\gamma_5 =  
  \begin{pmatrix}
  I&   0\\
  0&  -I
  \end{pmatrix}.
  \enq 
The matrix $\ga_5$ is also hermitian and anti commutes with $\al$ and $\be$
\begin{eqnarray} \label{DM2}
\ga_5^\dagger = \ga_5,~~~ \ga_5^2 = 1,\\ 
\{\ga_5, \al\} = \{\ga_5, \be\} = 0.
\end{eqnarray}

The SUSY  LF Hamiltonian $H_{LF}$ is given by the $sl(1/1)$ algebra
\beqa
\{Q,Q^\dagger\} &=& H_{LF},  \\
\{Q,Q\} &=& \{Q^\dagger, Q^\dagger\} = 0,
\enqa
but the supercharges $Q$ and $Q^\dagger$ are now represented by $4 \times 4$ matrices. Furthermore,  since the Hamiltonian operator $H_{LF} = P_\mu P^\mu  = M^2$ is invariant, it implies that $H_{LF}$ should  depend on a frame independent variable.  In  impact space the relevant invariant variable is $\zeta$, and thus the representation:
\beq
 Q =   \eta  \left(\frac{d}{d \zeta}  + W(\zeta)  \right),
 \enq
 and
 \beq
 Q^\dagger =   \eta^\dagger \left(- \frac{d}{d \zeta} + W(\zeta)\right),
 \enq
 where the spinor operators  $\eta$ and $\eta^\dagger$ satisfy the anti-commutation relation
 \beq
 \{\eta, \eta^\dagger\} = 1,
 \enq
and are given by 
\beq
\eta  = \frac{\be + i  \al}{2},  \quad  \eta^{\dagger} = \frac{\be - i \al}{2},
\enq
in the  $4 \times 4$ matrix representation defined above. We also have
\beq   \label{ga5}
[\eta, \eta^\dagger] = \ga_5.
\enq
The LF Hamiltonian is thus expressed as
\beq \label{HLF}
H_{LF} =  \{Q, Q^\dagger\} =  - \frac{d^2}{d \zeta^2}  + W^2(\zeta) +   \ga_5 \, W'(\zeta),
\enq
which is frame independent.

\subsection{A Linear Dirac Equation from Supersymmetric Quantum Mechanics in the Light-Front}

Since $\ga_5^2 = 1$, the LF Hamiltonian \req{HLF} can  be conveniently expressed as $H_{LF} =  B B^\dagger$ where
\beq \label{B1}
B =    \left(\frac{d}{d \zeta} + \ga_5 \, W(\zeta) \right),
 \enq
 and
 \beq  \label{B2}
 B^\dagger =  \left(- \frac{d}{d \zeta} + \ga_5 W(\zeta)\right).
\enq

The next step is to take the `square root' of the Hamiltonian $H_{LF}$. For this purpose we write $H_{LF}$ as a product  of Hermitian operators which we label $D_{LF}$; thus 
$H_{LF} = D_{LF}^2$. Using the relation $i \alpha B = - i B^\dagger \al$ and equations \req{B1} and \req{B2}, we have
\beq \label{DEzeta}
D_{LF} =  -i \al   \left(-\frac{d}{d \zeta} + \ga_5 \, W(\zeta) \right),
 \enq
 and thus the invariant Dirac equation~\cite{Brodsky:2008pg}
 \begin{equation} \label{DE}
\left( D_{LF} - M \right) \psi(\zeta) = 0,
\end{equation}
where $\psi(\zeta)$ is a LF Dirac spinor.
Premultiplying the linear Dirac wave equation (\ref{DE}) by the operator $D_{LF} + M$ and using the
properties of the Dirac matrices given above, we recover the LF eigenvalue equation
\begin{equation} \label{DLFH}
H_{LF}  \,  \psi = D_{LF}^2 \psi = M^2  \psi,
\end{equation}
where  $H_{LF}$ is given by \req{HLF}. We thus reproduce the results obtained in Ref. \cite{Brodsky:2008pg} using an operator construction of the light-front Hamiltonian and the Dirac equation,  but starting from light-front supersymmetric quantum mechanics~\cite{FootnoteDirac}.

It is convenient to separate the kinematic and dynamic contributions to the superpotential. We write
\beq  \label{LFSP}
W(\zeta) = \frac{\nu + 1/2}{\zeta} + u(\zeta),
\enq
where $\nu$ is a dimensionless parameter representing the LF orbital angular momentum, and  the dynamical effects are encoded in the function $u(\zeta)$. From \req {DEzeta} we can express the LF-invariant Dirac equation \req{DE} for the superpotential \req{LFSP}  as a system of coupled linear differential equations
\begin{eqnarray} \label{LFDEmtrx}  \nonumber
- \frac{d}{d \zeta} \psi_-  - \frac{\nu+\half}{\zeta}\psi_-  -  u(\zeta) \psi_-&=& M \psi_+ , \\
 \frac{d}{d \zeta} \psi_+ - \frac{\nu+\half}{\zeta}\psi_+  - u(\zeta) \psi_+ &=& M \psi_- , 
\end{eqnarray}
where  the chiral spinors are defined by $\psi_\pm = \half \left( 1 \pm \ga_5 \right) \psi.$

\subsection{Holographic Embedding}

We can now determine the  LF superpotential  $u(\zeta)$ in \req{LFSP} for arbitrary half-integer spin by embedding the LF results in AdS space.  We start with an effective action for  Rarita-Schwinger (RS) spinors in AdS space $\left[ \Psi_{N_1 \cdots N_T}\right]_\al$, which transform as symmetric tensors of rank $T$ with indices $N_1 \dots N_T$, and as  Dirac spinors with index $\alpha$~\cite{Rarita:1941mf}. In presence of an effective interaction $V(z)$ the effective action is given by~\cite{deTeramond:2013it}
 \begin{multline} \label{af}
~~~  S_{\it  eff} = \half  \int  \!
d^{d} x \,dz\,  \sqrt{g} \, g^{N_1\,N_1'} \cdots g^{N_T\,N_T'}  \\
\left[ \bar  \Psi_{N_1 \cdots N_T}  \Big( i \, \Ga^A\, e^M_A\, D_M
-  \mu - V(z)\Big)
 \Psi_{N_1' \cdots N_T'} + h.c. \right] , ~~~~~
 \end{multline} 
where $\sqrt{g} = \left(\frac{R}{z}\right)^{d+1}$ and $e^M_A$ is the inverse vielbein, $e^M_A = \left(\frac{z}{R}\right)
\delta^M_A$.  The covariant derivative $D_M$  includes the affine connection and the spin connection. The tangent-space Dirac matrices obey the usual anti-commutation relation $\left\{\Gamma^A, \Gamma^B\right\} = 2 \eta^{A B}$.  We have not included a dilaton factor  $e^{\vp(z)}$ in \req{af} since it 
can be  absorbed by redefining  the RS spinor according to $ \Psi_T \to e^{\vp(z)/2} \Psi_T$~\cite{Kirsch:2006he, deTeramond:2013it}.   This is a consequence of the linear covariant derivatives in the fermion action, which also prevents a mixing between dynamical and kinematical effects, and thus, in contrast to the effective action for integer spin fields \req{action2}, the AdS mass $\mu$  in Eq. \req{af} is constant. Since a dilaton factor has no dynamical consequences, one must introduce an effective confining interaction $V(z)$ in the fermion action to break conformal symmetry and generate a baryon spectrum~\cite{Brodsky:2008pg, Abidin:2009hr}.

It is shown below that the potential $V(z)$, which has been introduced hitherto {\it ad hoc}, is precisely related to the superpotential $u$ \req{LFSP}. Furthermore, in Sec. \ref{LFSCQM} it is shown that, in analogy with the boson case~\cite{Brodsky:2013ar}, the form of $u$ is determined in the framework of the superconformal algebra.

A physical baryon has plane-wave solutions  with four-momentum $P_\mu$,  invariant mass $P_\mu P^\mu = M^2$, and polarization indices along the physical coordinates. Factoring out  the four-dimensional plane-wave and spinor dependence, as well as the scale factor $(1/z)^{T-d/2}$, we have
\beq \label{Psipsi}
\Psi^\pm_{\nu_1 \cdots \nu_T}(z)   = e^{ i P \cdot x}  \,  u^\pm_{\nu_1 \cdots \nu_T} ({P}) \, \left(\frac{R}{z} \right)^{T-d/2}   \Psi^\pm_T(z),
\enq
where $T = J - \half$ and the fully symmetric RS chiral spinor  $u^\pm_{\nu_1 \dots \nu_T} =
\half (1 \pm \gamma_5)u_{\nu_1 \dots \nu_T}$  satisfies the four-dimensional chirality equations
 \beq
 \gamma \cdot P \, u^\pm_{\nu_1 \dots \nu_T}({P}) = M  u^\mp_{\nu_1 \dots \nu_T}({P}), \qquad
\gamma_5 u^\pm_{\nu_1 \dots \nu_T}({P}) = \pm \, u^\pm_{\nu_1 \dots \nu_T}({P}).
\enq
Variation of the AdS action \req{af} leads for $d =4$ to the Dirac equation
\begin{eqnarray} \label{AdSDEmtrx}  \nonumber
- \frac{d}{d z} \Psi^-_T   - \frac{\mu R}{z} \Psi^-_T  -  \frac{R}{z} \, V(z)   \Psi^-_T &=& M \Psi^+_T , \\
   \frac{d}{d z}  \Psi^+_T - \frac{\mu R}{z} \Psi^+_T  - \frac{R}{z} \, V(z)   \Psi^+_T &=& M \Psi^-_T , 
\end{eqnarray}
and  the Rarita-Schwinger condition~\cite{Rarita:1941mf}  in physical space-time~\cite{deTeramond:2013it}
 \beq \label{dirac-SE1}
 \gamma^\nu \Psi_{\nu  \nu_2 \, \dots \,\nu_T} =0.
 \enq

By identifying the holographic variable $z$ with the invariant LF variable $\zeta$ and the AdS  LF spinors  by the mapping  $\Psi_T^\pm(z) \to \psi^\pm(\zeta)$, we can  compare  \req{AdSDEmtrx} with \req{LFDEmtrx}. Provided that the AdS mass $\mu$  is related to the parameter $\nu$ by 
\beq  \label{munuhalf}
\mu R =  \nu + \half,
\enq
the specific LF mapping  gives a relation between  the effective interaction $V(z)$ in the AdS action \req{af} and the function $u(\zeta)$ in the LF superpotential \req{LFSP} 
\beq
u(\zeta) = \frac{R}{\zeta} \, V(\zeta).
\enq
In fact they are identical (modulo a kinematic factor), and this relation thus leads to a $J$-independent potential.   This is a remarkable result, since independently of the specific form of the potential, the value of the baryon masses along a given Regge trajectory depends only on the LF orbital angular momentum $L$~\cite{Gutsche}. Thus, in contrast with the vector mesons \req{UJ}, there is no spin-orbit coupling, in agreement with the observed near-degeneracy in the baryon spectrum~\cite{Klempt:2009pi, Footnote3}.

\section{Light-Front Superconformal Quantum Mechanics \label{LFSCQM}}

In order to fix the superpotential $u$ \req{LFSP} we follow  Fubini and Rabinovici in Ref.~\cite{Fubini:1984hf}, and consider a one-dimensional quantum field theory invariant under  conformal  and supersymmetric transformations. Imposing  conformal symmetry leads to a unique choice of the superpotential $W$ \req{Q}, namely
\beq \label{f}
W(x) =  \frac{ f }{x},
\enq
in order for $f$ to be a dimensionless constant. In this case the graded-Lie algebra has, in addition to the Hamiltonian $H$ and the supercharges $Q$ and $Q^\dagger$, an additional generator $S$ which is the square root  of the generator of conformal transformations $K$.  The enlarged algebraic structure is the superconformal algebra of Haag, Lopuszanski and Sohnius~\cite{Fubini:1984hf, Haag:1974qh, refM}.  Using the one-dimensional quantum-mechanical representation of the operators
\beqa
 Q   &=&    \chi \left(\frac{d}{d x} + \frac{f}{x} \right), \\
 Q^\dagger &=& \chi^\dagger \left(- \frac{d}{d x} + \frac{f}{x}\right),  \\
 S &=& \chi \, x,\\
  S^\dagger &=& \chi^\dagger x,
\enqa
 it is simple to verify that the algebraic structure of the enlarged algebra is fulfilled. We find
\beqa
\half \{Q,Q^\dagger\} &=& H,  \quad  \quad \half \{S,S^\dagger\} = K, \\
\half \{Q ,S^\dagger\} &=& \frac{f}{2} + \frac{\si_3}{4} -  D,\\
\half \{Q^\dagger ,S\} &=& \frac{f}{2} + \frac{\si_3}{4} +  D,
\enqa
where the operators
\beqa
H & = & \frac{1}{2} \left( - \frac{d^2}{d x^2}  + \frac{f^2 -  \si_3 f}{x^2} \right), \\
 K  & = & \half x^2, \\
 D & = & \frac{i}{4} \left(\frac{d}{dx} x + x \frac{d}{d x} \right).
\enqa
satisfy the conformal algebra \req{CA}. The anticommutation of all other generators vanish:  $\{Q,Q\} = \{Q^\dagger, Q^\dagger\} =  \{Q,S\} = \cdots = 0$.

In analogy with the dAFF procedure~\cite{deAlfaro:1976je}, we now define, following  Fubini and Rabinovici~§\cite{Fubini:1984hf}, a new supercharge $R$ as a linear combination of the generators $Q$ and $S$~
\beq
R =  \sqrt{u} \, Q + \sqrt{w} \, S,
\enq
and compute a new Hamiltonian $G$
\beq
G = \half\{R, R^\dagger\}.
\enq
We find
\beq \label{Gwu}
G = u  H + w  K + \half \sqrt{u w} \,(2 f + \si_3),
\enq
which is a compact operator for   $u w >0$.

The  quantum mechanical evolution operator $G$ \req{Gwu}  obtained by this procedure  is analogous to the Hamiltonian \req{HtauQ} obtained by the procedure of de Alfaro, Fubini and Furlan~\cite{deAlfaro:1976je}. Remarkably, in the superconformal case there appears beside the confining term $w \,K$ also  a constant term $\half \sqrt{u w} (2f \pm 1)$ in $G$, which, as we will describe below, plays a key role in explaining the correct phenomenology.

\subsection{Superconformal Quantum Mechanics in the Light-Front \label{LFSCQMA}}

The light-front extension of the superconformal results follows  from  the LF superpotential
\beq \label{LFCSP}
W(\zeta) = \frac{\nu + 1/2}{\zeta},
\enq
which corresponds to a kinematic term in the LF Hamiltonian.   We now extend the new Hamiltonian $G$  \req{Gwu} to a relativistic LF Hamiltonian by the method described in Sec. \ref{LFSUSYQMA}.  This amounts  to replace the Pauli matrix $\si_3$ in  \req{Gwu}  with $\ga_5$ in \req{ga5}.  We obtain:
\beqa\label{eq:LFHs}
H_{LF}  &=&  \{R, R^\dagger\} \nn \\
& = & - \frac{d^2}{d \zeta^2} 
+ \frac{\left(\nu + \half\right)^2}{\zeta^2} - \frac{\nu +
  \half}{\zeta^2} \gamma_5 + \la_B^2 \zeta^2 + \la_B (2 \nu + 1) + \la_B \gamma_5,
\enqa
where the arbitrary  coefficients $u$ and $w$ in \req{Gwu} are fixed to   $u = 1$ and $w =  \la_B^2$.  Thus the supercharge $R$ is the superposition
\beq
R = Q + \la_B S  \label{R} .
\enq

In $2 \times 2$ block-matrix form the light-front Hamiltonian \req{eq:LFHs} can be expressed as

 \beq \label{HLFu}
H_{LF} =  \left(  \! \begin{array}{cc}
    - \frac{d^2}{d \zeta^2}  - \frac{1 - 4 \nu^2}{4 \zeta^2} +  \la_B^2 \zeta^2 +
2 \la_B (\nu + 1) & 0\\ 
    0 & - \frac{d^2}{d \zeta^2} - \frac{1 - 4(\nu + 1)^2}{4 \zeta^2} + \la_B^2 \zeta^2 +
2 \la_B  \nu   \\  
  \end{array} \! \right).
  \enq
The light-front eigenvalue equation $H_{LF} \vert \psi \rangle = M^2 \vert \psi \rangle$ has eigenfunctions
\begin{eqnarray} \label{psip}
\psi_+(\zeta) &\sim& \zeta^{\frac{1}{2} + \nu} e^{-\la_B \zeta^2/2}
  L_n^\nu(\la_B \zeta^2) ,\\ \label{psim}
\psi_-(\zeta) &\sim&  \zeta^{\frac{3}{2} + \nu} e^{-\la_B \zeta^2/2}
 L_n^{\nu+1}(\la_B \zeta^2), 
\end{eqnarray}
and  eigenvalues
\begin{equation}  \label{M2F}
M^2 = 4 \la_B (n + \nu + 1).
\end{equation}
As a consequence of parity invariance, the eigenvalues for the chirality plus and minus eigenfunctions are identical. One can also show that the probabilities for both  components $\psi_+$  and $\psi_-$ are the same (See appendix \ref{appendix})
\begin{equation}  \label{Ppm}
\int d\zeta \,  \psi^2 _+(\zeta)   =  \int d \zeta  \, \psi^2_-(\zeta)  .
\end{equation}
For $\la_B <0$ no solution is possible.

\section{Systematics of the Baryon Spectrum \label{BS}}

To determine how well the superconformal light-front holographic model encompasses the systematics of the baryon spectrum,  we list in Table \ref{baryons} the confirmed (3-star and 4-star) baryon states from the Particle Data Group~\cite{Agashe:2014kda}. The internal spin, light-front internal orbital angular momentum and radial quantum number assignment of the $N$ and $\Delta$ excitation spectrum is found from the total angular momentum-parity PDG assignment using the conventional $SU(6) \supset SU(3)_{flavor} \times SU(2)_{spin}$ multiplet structure~\cite{Lichtenberg:1978pc}, but other model choices are also possible~\cite{Footnote4}. Further details can be found in \cite{Brodsky:2014yha}.

\begin{table}[ht]
 \caption{\label{baryons}\small Classification of confirmed baryons listed by the PDG~\cite{Agashe:2014kda}. The labels $L$,  $S$ and  $n$ refer to the internal  orbital angular momentum, internal spin and radial quantum number respectively. The even-parity baryons correspond to the $\bf 56$ multiplet of $SU(6)$ and the odd-parity to the ${\bf 70}$.}
 {\begin{tabular}{@{}ccccc@{}}
 \hline\hline 
  $SU(6)$ &  $~S~$ & $ L$ &  $~n~$ &   Baryon State \\[0.0ex]
 \hline
 \multicolumn{5}{c}{}\\[-7.0ex]
 ${\bf 56}$  & $\half$ & 0  &  0  & $N{\half^+}(940)$\\[-1.0ex]
 {}    &   $\threehalf$& 0 &   0  &$\Delta{\threehalf^+}(1232)$ \\[0.0ex]
 \hline  \\[-7.0ex]
  ${\bf 56} $     &   $\half$ & 0  &  1  & $N{\half^+}(1440)$\\[-1.0ex]
  {}  &   $\threehalf$& 0 &   1  &$\Delta{\threehalf^+}(1600)$\\[0.0ex]   
  \hline\\[-7.0ex]
 ${\bf 70}$ & $\half$    & 1 & 0  & $N{\half^-}(1535)~~ N{\threehalf^-}(1520)$ \\[-1.0ex]
 {}  & $\threehalf$ & 1 &  0  & $N{\half^-}(1650)~~ N{\threehalf^-}(1700)~~N{\fivehalf^-}(1675)$\\[-1.0ex]
 {}  & $\half$ & 1  &  0  &$\Delta{\half^-}(1620)~~ \Delta{\threehalf^-}(1700)$ \\[0.0ex]
 \hline  \\[-7.0ex]
  ${\bf 56}$      &  $\half$ & 0  &  2  & $N{\half^+}(1710)$\\[-1.0ex]
 ${}$  & $\half$ & 2  &  0  &$N{\threehalf^+}(1720)~~ N{\fivehalf^+}(1680)$ \\[-1.0ex]
 {}    & $\threehalf$ & 2  & 0  &~ $\Delta{\half^+}(1910)~~ \Delta{\threehalf^+}(1920)
                     ~~ \Delta{\fivehalf^+}(1905)~~\Delta{\sevenhalf^+}(1950)$\\[0.0ex]
 \hline \\[-7.0ex]
{\bf 70}   & $\threehalf$ & 1 &  1  & $N{\half^-}~~~~~~~~ ~~ N{\threehalf^-}(1875)~~N{\fivehalf^-}~~~~~~~~$\\[0.0ex]
 \hline \\[-7.0ex]
  ${}$      & $\threehalf$ & 1  &  1  &$\Delta{\fivehalf^-(1930)} $  \\[0.0ex]
 \hline   \\[-7.0ex]
  ${\bf 56}$      & $\half$ & 2  &  1  &$N{\threehalf^+(1900)}  ~~ N{\fivehalf^+}$~~~~~~~~ \\[0.0ex]
   \hline \\[-7.0ex]
 ${\bf 70}$ &    $\half$ & 3 &  0  & $N{\fivehalf^-}~~ ~~~~~~ N{\sevenhalf^-}$ \\[-1.0ex]
 {}   &$\threehalf$ & 3  & 0  &   $N{\threehalf^-}~~~~~~ ~~ N{\fivehalf^-}~~~~~~ ~~
                     N{\sevenhalf^-}(2190)~~ N{\ninehalf^-}(2250)$\\[-1.0ex]
 {}  &   $\half$ & 3  & 0   & $\Delta{\fivehalf^-}~~~~~ ~~ \Delta{\sevenhalf^-}$ \\[0.0ex]
 \hline   \\[-7.0ex]
 ${\bf 56}$ &    $\half$ & 4 &  0  & $N{\sevenhalf^+}~~~~~~ ~~ N{\ninehalf^+}(2220)$ \\[-1.0ex]
 {}    &$\threehalf $ & 4 &  0  & $\Delta{\fivehalf^+}~~~~~~ ~~ \Delta{\sevenhalf^+} ~~~~~~ ~~
                       \Delta{\ninehalf^+}~~~~~~ ~~\Delta{\elevenhalf^+}(2420)$ \\[0.0ex]                      
  \hline   \\[-7.0ex]
 ${\bf 70}$   &  $\half$ & 5 &  0  &$N{\ninehalf^-}~~~~~~ ~~ N{\elevenhalf^-}~~~~~~~$ \\[-1.0ex]
 {}     &   $\threehalf$ & 5 &  0  & $N{\sevenhalf^-}~~~~~~~ ~~ N{\ninehalf^-}~~~~~~~
      N{\elevenhalf^-}(2600) ~~N{\thirteenhalf^-} $\\[0.0ex]
  \hline\hline
 \end{tabular}}
 \end{table}

The lowest possible stable state, the nucleon $N{\half^+}(940)$, corresponds to $n=0$ and $\nu = 0$. This fixes the scale $\sqrt \la_B =  M_P/2$. The resulting predictions for the spectroscopy of the positive-parity spin-$\half$ light nucleons are shown in  Fig. \ref{baryonspec} (a) for  the parent Regge trajectory for $n =0$ and  $\nu = 0, 2, 4, \cdots,  L $, where $L$ is the relative LF angular momentum between the active quark and the spectator cluster. Thus the dimensionless constant $f$ in the superpotential \req{f} is $f = L + \half$ for the plus parity nucleon trayectory.  The predictions for the daughter trajectories for $n=1$, $n = 2, \cdots $ are also shown in this figure. Only confirmed PDG~\cite{Agashe:2014kda} states are shown. The Roper state $N{\half^+}(1440)$ and the $N{\half^+}(1710)$ are well described  in this model as the first and second radial excited states of the nucleon. The newly identified state, the $N{\threehalf^+(1900)}$~\cite{Agashe:2014kda} is depicted here as the first radial excitation of the $N{\threehalf^+}(1720)$. The model is successful in explaining the $J$-degeneracy for states with the same orbital angular momentum observed in the light baryon spectrum, such as the $L = 2$ plus parity doublet  $N{\threehalf^+}(1720) - N{\fivehalf^+}(1680)$, which corresponds to and $J= \threehalf$ and $\fivehalf$ respectively (See Fig. \ref{baryonspec} (a)).

\begin{figure}[htdp]
\centering
\includegraphics[width=14.6cm]{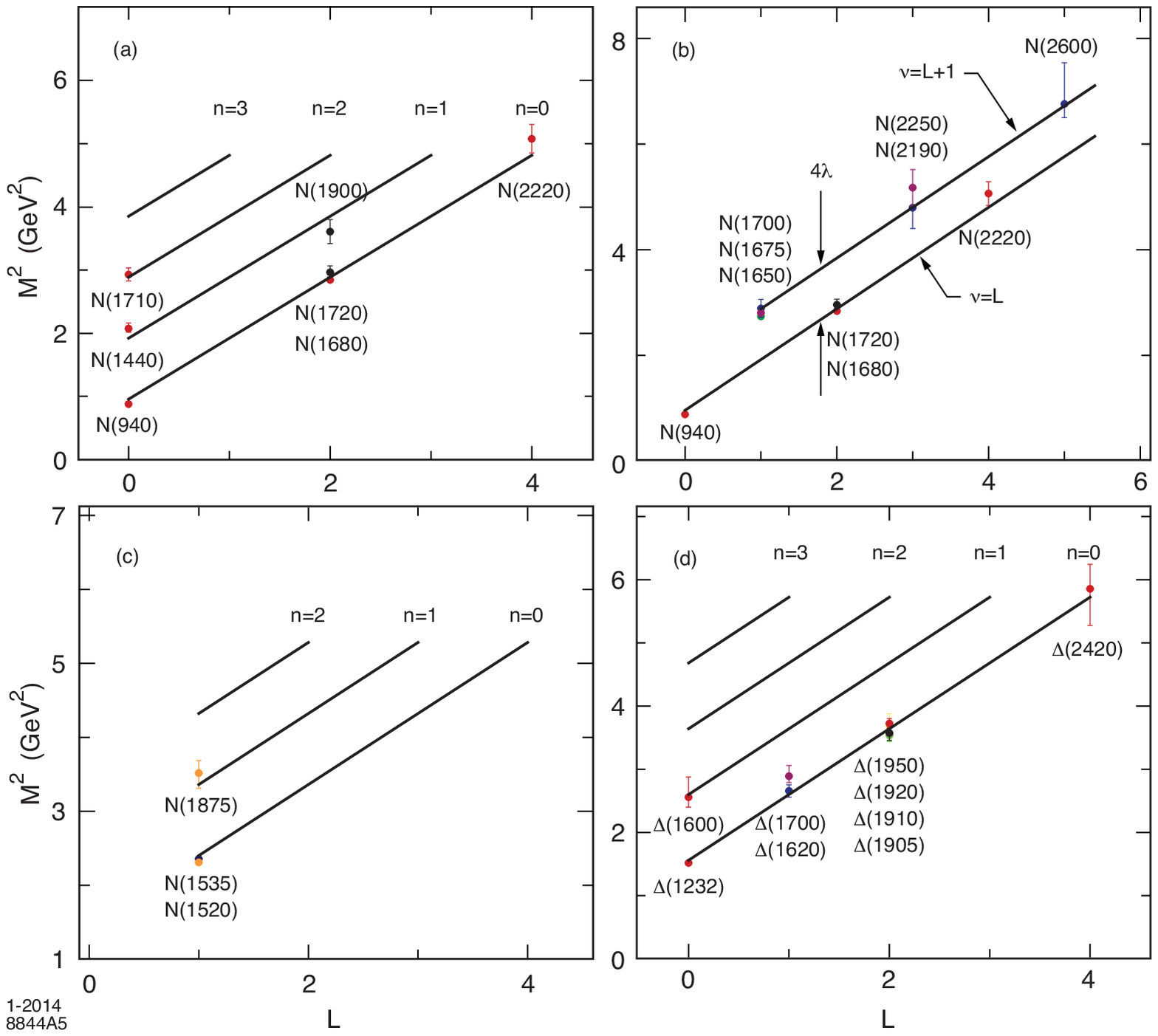}  
\caption{\small Orbital and radial baryon excitation spectrum.  Positive-parity spin-$\half$ nucleons (a) and spectrum gap between the negative-parity spin-$\threehalf$ and the positive-parity spin-$\half$ nucleons families (b). Minus parity spin-$\half$ $N$ ({c}) and plus and minus parity spin-$\half$ and spin-$\threehalf$ $\Delta$  families (d).  We have used  in this figure the value
$\sqrt{\la_B} = 0.49$ GeV for nucleons and  0.51 GeV for the Deltas.}
 \label{baryonspec}
\end{figure} 

In Fig. \ref{baryonspec} (b) we compare the positive parity  spin-$\half$ parent nucleon trajectory with the negative parity spin-$\threehalf$ nucleon trajectory.  As it is shown in this figure, the gap scale $4 \la$ determines not only the slope of the nucleon trajectories, but also the spectrum gap between the plus-parity spin-$\half$  and the minus-parity spin-$\threehalf$ nucleon families, as indicated by arrows in this figure. This means the respective assignment $\nu = L$ and $\nu = L+1$ for the lower and upper trajectories in Fig. \ref{baryonspec} (b), or  $f = L + \half$ and $f = L + \threehalf$ respectively. The degeneracy of states with the same orbital quantum number $L$ is also well described, as for example the degeneracy of the $L=1$ minus-parity triplet  $N{\half^-}(1650)$, $N{\threehalf^-}(1700)$, and $N{\fivehalf^-}(1675)$, which corresponds respectively to $J = \half$, $\threehalf$ and $\fivehalf$ (See: Fig. \ref{baryonspec} (b)).

Baryons  with negative parity and internal spin $S = \half$, such as the $N{\half^-}(1535)$, as well as baryon states with positive parity and internal spin $S = \threehalf$, such as the $\Delta{\threehalf^+}(1232)$ are well described by the assignment $\nu = L + \half$, or $f = L + 1$. This means, for example, that the positive and negative-parity $\Delta$ states are in the same trajectory consistent with experimental observations, as depicted in Fig. \ref{baryonspec} (d).  The newly found state, the $N{\threehalf^-}(1875)$~\cite{Agashe:2014kda}, depicted in Fig. \ref{baryonspec} ({c}) is well accounted as the first radial excitation of the $N{\threehalf^-}(1520)$. The degeneracy of the  $L=1$ minus-parity doublet  $N{\half^-}(1535) - N{\threehalf^-}(1520)$  for  $J = \half$ and $\threehalf$ is also well described. Likewise, the  $\Delta(1600)$ corresponds to the first radial excitation of the $\Delta(1232)$ as shown in Fig. \ref{baryonspec} (d).  The model explains  the degeneracy of the $L=2$ plus-parity quartet  $\Delta{\half^+}(1910)$, $\Delta{\threehalf^+}(1920)$, $\Delta{\fivehalf^+}(1905)$, and $\Delta{\sevenhalf^+}(1950)$ which corresponds  to $J = \half$, $\threehalf$, $\fivehalf$ and $\sevenhalf$ respectively (See: Fig. \ref{baryonspec} (d)). Our results for the $\Delta$ states agree with those of Ref.~\cite{Forkel:2007cm}. ``Chiral partners" such as the $N{\half^+}(940)$ and $N{\half^-}(1535)$ nucleons with the same total angular momentum $J = \half$, but with different orbital angular momentum and parity are non-degenerate from the onset.  To recapitulate, the parameter $f$,  has the internal spin $S$ and parity $P$ assignment  given in  Table \ref{nuasig}, which is equivalent to the assignment given in~\cite{deTeramond:2014yga}.

\begin{table}[ht] \label{nuT}
\caption{\label{nuasig}
\small Orbital quantum number assignment for the superpotential parameter $f$ for baryon trajectories  according to parity $P$ and internal spin $S$.}
\vspace{5pt}
\resizebox{6.0cm}{!}
 {
\begin{tabular}{ l | c r } 
\hline\hline
 & $~~S = \half$ & \, $~~~~ S = \threehalf$ \ ~~\\ [0.0ex]
 \hline
P = + ~& $~~f = L + \half$ &  $ f = L + 1$ \\ [-1.0ex]
 P = \ -- ~ &  $~ f = L + 1$ &  $f = L+ \threehalf$ \\ [0.0ex]
 \hline\hline
\end{tabular}
}
\end{table}

This particular assignment successfully describes the full light baryon orbital and radial excitation spectrum, and in particular the gap between trajectories with different parity and internal spin~\cite{deTeramond:2014yga}.   The assignment $\nu = L$ for the lowest trajectory, the proton trajectory, is straightforward and follows from the stability of the ground state, the proton, and the mapping of AdS to light-front physics. The assignment for other spin and parity baryons states, given in Table \ref{nuasig}, is motivated by the observed spectrum. It is hoped that further analysis of the different quark configurations and symmetries of the baryon wave function~\cite{Klempt:2009pi, Forkel:2008un, Selem:2006nd} will indeed explain the assignment of the dimensionless parameter $f$.

 If we follow the non-$SU(6)$ quantum number assignment for the  $\Delta{\fivehalf^-(1930)} $ given in Ref.~\cite{Klempt:2009pi}, namely  $S = 3/2$, $L =1$, $n=1$,  we find  with the present model the value $M_{\Delta(1930)} = 4 \sqrt \la_B = 2 M_P$, also consistent with the experimental result 1.96 GeV~\cite{Agashe:2014kda}.  An important feature of light-front holography and supersymmetric LF quantum mechanics is the fact that it predicts a similar multiplicity of states for mesons and baryons,  consistent with
 experimental observations~\cite{Klempt:2009pi}. This property is consistent with  the LF cluster decomposition of the holographic variable $\zeta$, which describes a system of partons as an active quark plus a cluster of $n-1$ spectators~\cite{Footnote3}. From this perspective, a baryon with 3  quarks looks in light-front holography as a quark--diquark system.

Another interesting consequence of the supersymmetric relation between the plus and minus chirality states, is the equal equal probability expressed by \req{Ppm}. This remarkable equality means that in the light-front holographic approach described here the proton's spin $J^z = L^z + S^z$ is carried by the quark orbital angular momentum:  $\langle J^z \rangle = \langle L^z_q \rangle = \pm 1/2$ since  $\langle S^z_q \rangle = 0$.

\section{Conclusions and Outlook \label{conclusions}}

In this article we have shown how superconformal quantum mechanics~\cite{Akulov:1984uh, Fubini:1984hf} can be extended to the light-front and how it can be precisely mapped to holographic QCD.    We have also examined the higher half-integer spin representations of the model by embedding the resulting Dirac invariant light-front wave equation in AdS space.  This procedure introduces a scale in the Hamiltonian equations and completely fixes the light-front potential in the Dirac equation introduced in Refs. \cite{Brodsky:2008pg, Abidin:2009hr}. In this approach the main features of the observed light-baryon spectrum are described.

The construction procedure is  similar to that of bosons~\cite{deAlfaro:1976je, Brodsky:2013ar}. Both are based on the breaking of conformal invariance within the algebraic structure, by a redefinition of the quantum mechanical evolution in terms of a superposition of the operators of the conformal or superconformal algebras. Since the generators have different dimensions this amounts to the introduction of a scale in the Hamiltonian while maintaining a conformal action. Compared with the holographic construction for baryons, this unified approach is more satisfactory.  In contrast to  the meson case,  the dilaton in the fermion action has no effect on the baryon spectrum.  Consequently,  a Yukawa potential must be introduced by hand to break conformal invariance.  Here, the same underlying principle is used to introduce a mass scale and generate the masses  for mesons and baryons from a spectrum generating algebra.  For baryons the quantum mechanical evolution is determined  from a supercharge which is  a superposition of elements of the superconformal algebra~\cite{Fubini:1984hf}.  In fact, the introduction of the generator $S$  (the square root of the generator of conformal transformations $K$) is the key step for extending the dAFF~\cite{deAlfaro:1976je, Brodsky:2013ar} procedure for obtaining a confining potential in the LF Dirac equation for baryons.

Mapping the results to  light-front bound-state equations  leads to a linear potential in the light-front Dirac equation and to a harmonic potential with additional constants in the quadratic Hamiltonian for fermions.  In contrast to the case of mesons,   there is no possibility to shift the energy levels by adding a constant to the linear potential in the light-front Dirac equation. Therefore superconformal quantum mechanics, together with the introduction of the scale according to Fubini and Rabinovici~\cite{Fubini:1984hf}, fixes completely the fermionic Hamiltonian. The equations of motion obtained  by following this procedure  are equivalent to the holographic light-front equations obtained from the fermion Lagrangian in AdS$_5$, with a Yukawa coupling providing  the effective potential.  In the bosonic case light-front holographic QCD yields  a $J$-dependent constant from the holographic embedding -- in addition to the confining harmonic potential obtained from conformal quantum mechanics~\cite{deAlfaro:1976je} -- which leads to a $J$-dependent level shift~\cite{Brodsky:2013ar}. Such  a  level shift is not possible for fermions,  and therefore there is a spin-$J$ degeneracy for states at fixed $L$ and $n$, an important characteristic which is actually observed in experiment.  The model is also consistent with similar Regge meson and baryon spectra and similar multiplicity of states for mesons and baryons.   In effect, the light-front Dirac equation for baryons described here is effectively a quark-diquark model. However, a quark-diquark construction is not imposed, but it is a natural consequence of the light-front cluster decomposition which follows from the LF embedding in AdS space~\cite{Footnote3}.  In this approach the quark and diquark are both massless.

In this paper we have described a mechanism for the emergence of a confining light-front Hamiltonian for hadrons. A mass scale $\sqrt \la$ and confining potentials appear in the light-front Schr\"odinger and Dirac equations, consistent with the conformal invariance of the action,  by applying the group-theoretical  methods of Refs.~\cite{deAlfaro:1976je, Fubini:1984hf}. We have given a relation between the dimensionless quantities $L$, $f$ or $g$, and $\mu R$ occurring in the light-front Hamiltonian, the quantum mechanical evolution operator in the algebraic approach, and  the wave equations in AdS$_5$, respectively (See Eqs. \req{gL}, \req{muJL}, \req{munuhalf} and Table \ref{nuasig}). We expect that further analysis of the different quark configurations and symmetries of the hadron wavefunctions will shed further light on the detailed relations  between  these dimensionless parameters.

Even if a supersymmetric connection inspired by the universality of the Regge trajectories for bosons and baryons was our starting point, in the context of this article the supersymmetric construction of baryonic states refers to the ``supersymmetry'' between positive and negative chirality of light-front spinors. In this case supersymmetry is not broken since  there is a perfect pairing for each baryonic state including the ground state,  consistent with parity invariance.  This does not exclude the possible supersymmetric connections between mesons and baryons which would be manifest as a consequence of confinement dynamics.  In fact, although the form of the potential is fixed in both cases by the dAFF procedure and its extension to the superconformal algebra, the numerical values of the confining scales are {\it a priori} not related. Nevertheless the values of $\lambda$ for the coefficient of the confining potentials come out to be similar in both cases with similar spacing between the orbital and radial hadronic excitations. This suggests a supersymmetric relation between the  underlying dynamics of the observed bosonic and fermionic hadrons. In this case, supersymmetry is  broken since the ground state, the pion,  is massless in the chiral limit and is not paired. We shall treat this  subject elsewhere.

\acknowledgments{We thank Vittorio de Alfaro,  Cedric Lorc\'e and Michael Peskin for helpful comments. This work is supported by the Department of Energy  contract DE--AC02--76SF00515.}

\appendix*

\section{Supercharges and Ladder Operators\label{appendix}}

The supercharge operator $R$ \req{R} in the light-front quantum mechanical representation discussed in Sect. \ref{LFSCQM} can be expressed as
\beqa
R &=& Q + \la S = \eta \, b,   \\
R^\dagger &=& Q^\dagger + \la S^\dagger = \eta^\dagger  b^\dagger ,
\enqa
where  the spinor operators $\eta$ and $\eta^\dagger$ in a $4 \times 4$ matrix representation are 
\beq 
 \eta =
  \begin{pmatrix}
  0 & I \\ 
  0 & 0
  \end{pmatrix}, ~~~~~
 \eta^\dagger =
  \begin{pmatrix}
  0 & 0\\
  I  & 0
  \end{pmatrix},~~~~~
\enq
with $I$ a two-dimensional unit matrix,  and the operators $b$ and $b^\dagger$ are given by
\beqa
b_\nu  &=&  \frac{d}{d \zeta} + \frac{\nu + \half}{\zeta} + \la \zeta,\\
b^\dagger_\nu &=&  - \frac{d}{d \zeta} + \frac{\nu + \half}{\zeta} + \la \zeta.
\enqa
The LF Hamiltonian $H_{LF}$ \req{HLFu} is conveniently factorized in terms of the linear operators $b$ 
\beq \label{HRR}
H^\nu_{LF} =   \{R, R^\dagger\} = \left(  \! \begin{array}{cc}
   b_\nu b^\dagger_\nu & 0\\ 
    0 & b_\nu^\dagger b_\nu  \\  
  \end{array} \! \right),
  \enq
and is thus integrable~\cite{Infeld:1941, Infeld:1951mw}.

Consider the eigenvalue equation for $b_\nu b_\nu^\dagger$
\begin{equation} \label{eq:LFWEx}
\left(- \frac{d^2}{d x^2} - \frac{1 - 4 \nu^2}{4 x^2} + \ka^2  \zeta^2 + 2 \ka(\nu + 1) \right)
\phi_\nu(x) = \phi_\nu(x) ,
\end{equation}
where $x = \zeta M$ and $\kappa = \la/M$.  Equation (\ref{eq:LFWEx}) is equivalent to
$b_\nu b_\nu^\dagger \vert \nu \rangle = \vert \nu \rangle $. It is also simple to verify that
$b_\nu^\dagger \vert \nu \rangle \sim  \vert \nu + 1 \rangle$
or
\begin{equation} \label{bdzeta}
\left(- \frac{d}{d \zeta} + \frac{\nu + \half}{\zeta } + \la \, \zeta \right) \phi_\nu(\zeta)  \sim
 \phi_{\nu+1}(\zeta).
\end{equation}
Likewise, one can show that  $b_\nu  \vert \nu \rangle  \sim  \vert \nu - 1 \rangle$.

We now construct a new supercharge $T$ and its adjoint $T^\dagger$ as the linear superposition~\cite{Fubini:1984hf}
\beqa
T &=& Q^\dagger - \la S^\dagger = \eta^\dagger  a,  \label{T}\\
T^\dagger &=& Q - \la S = \eta \, a^\dagger,    \label{Tdag}
\enqa
where
\beqa
a_\nu &=&  - \frac{d}{d \zeta} + \frac{\nu + \half}{\zeta} - \la \zeta, \label{a} \\
a^\dagger_\nu &=&   \frac{d}{d \zeta} + \frac{\nu + \half}{\zeta} - \la \zeta.
\enqa

One can show  that the operator \req{a}  lowers the radial quantum number $n$ by one unit and raises $\nu$ by  one unit
\begin{equation}
a \vert n, \nu \rangle \sim \vert n - 1, \nu + 1 \rangle.
\end{equation}
For a given $\nu$ the lowest possible state 
corresponds to $n = 0$. Consequently the state $\vert n=0, \nu \rangle$
is annihilated by the action of the operator $a$, $a \vert n=0, \nu\rangle = 0$,
or equivalently
\begin{equation}
\left(\frac{d}{d \zeta} - \frac{\nu + \half}{\zeta} 
+ \la \zeta\right) \phi^{n=0}_\nu(\zeta) =0,
\end{equation}
with solution
\begin{equation}
\phi^{n=0}_\nu(\zeta) = C_\nu \zeta^{1/2 + \nu} e^{-\la \zeta^2/2},
\end{equation}
where $C_\nu$ is a constant. Writing
\begin{equation}
\phi_\nu(\zeta) = C_\nu \zeta^{1/2 + \nu} e^{-\la \zeta^2/2} G_\nu(\zeta),
\end{equation}
and substituting in \req{bdzeta} we get
\begin{equation}
2 x \, G_\nu(x) - G_\nu '(x) \sim  x \, G_{\nu + 1}(x),
\end{equation}
with $x = \sqrt{\la} \, \zeta$, 
a relation which defines the confluent hypergeometric function $U(n, \nu + 1, x)$ in terms of $U(n, \nu, x)$~\cite{AS64}
\beq
U(n, \nu + 1, x) =  U(n, \nu, x)  - U'(n, \nu, x),
\enq
or equivalently
\beq
2 x \, U(n, \nu + 1, x^2) =  2 x \, U(n, \nu, x^2)  - \frac{d U'(n, \nu, x^2)}{dx}.
\enq
Thus the normalizable  solution to the eigenvalue equation  $b b^\dagger  \phi(\zeta) = M^2  \phi(\zeta)$:
\begin{equation}
\phi_{n, \nu}(\zeta) = C_\nu \zeta^{1/2 + \nu} e^{-\la \zeta^2/2}
L_n^\nu(\la \zeta^2).
\end{equation}
The  solution also follows  from the iterative  application of the ladder operators following the procedure described in~\cite{Arik:2008}.  We find
\beq
\phi(\zeta)_{n, \nu}  \sim  \zeta^{1/2 - \nu} e^{\la \zeta^2/2} \left(\frac{1}{\zeta} \frac{d}{d \zeta} \right)^n \zeta^{2(n + \nu)} e^{- \la \zeta^2},
\enq
with eigenvalues
\beq
M^2 = 4 \la (n + \nu + 1).
\enq

Since we know the general solution for the upper component  of the spinor wavefunction $\phi_\nu$, it is straightforward to compute the lowest component $b^\dagger \phi_\nu$ , with identical mass, by applying the supercharge operators. We find
\beqa
T \begin{pmatrix}
\phi_{n, \nu} \\  0
\end{pmatrix} &=& 0, \\
R^\dagger \begin{pmatrix}
\phi_{n, \nu} \\  0
\end{pmatrix} &=& 
C_{n, \nu} \begin{pmatrix}
 0 \\   \phi_{n, \nu + 1}
\end{pmatrix} ,
\enqa
with
\beq
C_{n,\nu} = \sqrt{\frac{\la}{n + \nu + 1}}.
\enq
Thus the solution
\beqa
\psi(\zeta) &=& \psi_+ u_+ + \psi_- u_- \\
&=& C z^{\frac{1}{2} + \nu} e^{-\la \zeta^2/2}
\left[ L_n^\nu\left(\la \zeta^2\right) u_+
   + \frac{\sqrt{\la} \zeta}{\sqrt{n +\nu + 1}}  
L_n^{\nu+1}\left(\la \zeta^2\right) u_- \right],
\enqa
with normalization
\begin{equation} \label{eq:spnorm}
\int d\zeta \,  \psi^2 _+(\zeta)   =  \int d \zeta  \, \psi^2_-(\zeta)  .
\end{equation}
Identical results follow by solving directly the Dirac equation  \req{LFDEmtrx} for the conformal superpotential \req{LFSP} with $u = \la \zeta$.

The light-front quantum mechanical evolution operator $H^\nu_{LF}$ \req{HRR} was constructed in terms of the supercharges $R$ and $R^\dagger$.  We can also construct a light-front Hamiltonian  $\overline{H}^\nu_{LF}$ in terms of the supercharges $T$ and $T^\dagger$  given by \req{T} and \req{Tdag}:
\beq \label{HTT}
\overline{H}^\nu_{LF} =   \{R, R^\dagger\} = \left(  \! \begin{array}{cc}
    a^\dagger_\nu a_\nu & 0\\ 
    0 &   a_\nu a_\nu^\dagger  \\  
  \end{array} \! \right).
  \enq
The light-front Hamiltonians $\overline{H}_{LF}$ \req{HTT} and $H_{LF}$ \req{HRR} are related by  $\overline{H}_{LF}(\la) = H_{LF}(-\la)$.



\begin{thebibliography}{00}
\bibitem{Footnote1}
 For a hadron with four-momentum  $P^\mu =  (P^+, P^-, \mbf{P}_{\!\perp})$, $P^\pm = P^0 \pm P^3$, the invariant light-front Hamiltonian is  $H_{LF}= P_\mu P^\mu = P^+ P^-  - \mbf{P}_\perp^2$, where $P^+$ and  $\mbf{P}_{\!\perp}$ are kinematical operators and the Hamiltonian $P^- = i \frac{\partial}{\partial x^+}$ generates the quantum evolution of the system with respect to the light-front time $x^+ = x^0 + x^+$.
\bibitem{Footnote2}
 For a two-parton bound state $\zeta = \sqrt{x(1-x)} \vert \mbf{b}_\perp \vert$, where $x$ is the longitudinal momentum fraction and $ \mbf{b}_\perp$ is  the transverse-impact distance between the two quarks and $L = {\rm max} \vert L^3 \vert$.
\end{thebibliography}
\end{document}